\documentclass[a4paper]{article}

\usepackage{INTERSPEECH2020}

\title{The IDLAB VoxCeleb Speaker Recognition Challenge 2020 System Description}
\name{Jenthe Thienpondt\thanks{$^*$Equal contribution}$^*$, Brecht Desplanques$^*$, Kris Demuynck}
\address{
    IDLab, Department of Electronics and Information Systems, Ghent University - imec, Belgium
}
\email{jenthe.thienpondt@ugent.be, brecht.desplanques@ugent.be}

\begin{document}

\maketitle
\begin{abstract}
In this technical report we describe the IDLAB top-scoring submissions for the VoxCeleb Speaker Recognition Challenge 2020 (VoxSRC-20) in the supervised and unsupervised speaker verification tracks. For the supervised verification tracks we trained 6 state-of-the-art ECAPA-TDNN systems and 4 Resnet34 based systems with architectural variations. On all models we apply a large margin fine-tuning strategy, which enables the training procedure to use higher margin penalties by using longer training utterances. In addition, we use quality-aware score calibration which introduces quality metrics in the calibration system to generate more consistent scores across varying levels of utterance conditions. A fusion of all systems with both enhancements applied led to the first place on the open and closed supervised verification tracks. The unsupervised system is trained through contrastive learning. Subsequent pseudo-label generation by iterative clustering of the training embeddings allows the use of supervised techniques. This procedure led to the winning submission on the unsupervised track, and its performance is closing in on supervised training. 
\end{abstract}

\section{Supervised speaker verification}

\begin{figure}[h]
  \centering
  \includegraphics[scale=0.2]{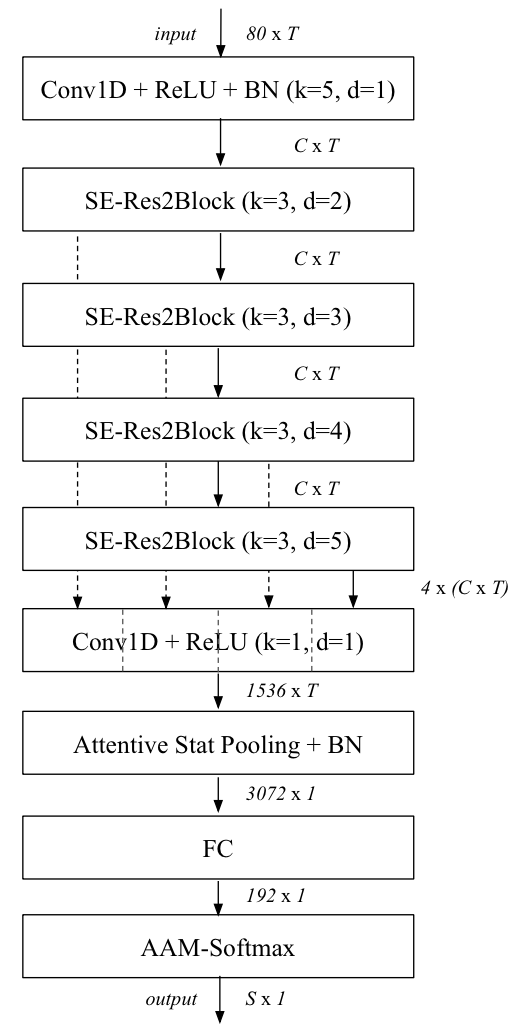}
  \setlength{\belowcaptionskip}{-10pt}
  \caption{ECAPA-TDNN system architecture. $T$ indicates the number of input frames, $C$ the amount of intermediate feature channels and $S$ the number of classification speakers. We denote \textit{k} and \textit{d} in the SE-Res2Block for the kernel size and dilation factor, respectively. See~\cite{ecapa_tdnn} for more details.}
  \label{fig:ecapa_tdnn}
\end{figure}

For the supervised verification track (VoxSRC-20 Track 1 \& 2) we train 6 systems based on our ECAPA-TDNN architecture~\cite{ecapa_tdnn, voxsrc20_submission} and 4 ResNet34~\cite{magneto} variants. The ECAPA-TDNN architecture is depicted in Figure~\ref{fig:ecapa_tdnn}. We scale up the network compared to~\cite{ecapa_tdnn} by using 2048 feature channels and add a fourth SE-Res2Block with a dilation factor of 5 for optimized verification performance for all models. To further diversify the models, we make small architectural changes across all five other ECAPA-TDNN models:

\begin{enumerate}
    \item Replacement of the ReLU-activation function in the middle layer of the attention module with the Tanh activation function.
    \item Addition of a Bi-directional Long Short-Term memory (BLSTM) layer in the first layer of the model~\cite{blstm_microsoft}.
    \item Increase of the embedding dimension from 192 to 256.
    \item 60-dimensional log Mel filter bank energies as input feature.
    \item Incorporation of two Sub-Centers per class in the AAM-softmax layer~\cite{subcenter} (SC-AAM), along with the integration of the dilation factor variability across the groups in the Res2Net grouped convolutions instead of across the SE-Res2Blocks. We refer to this as Dynamic Dilation (DD). Except for the first directly connected group, the 7 remaining groups in the central Res2Net~\cite{res2net} convolutions have dilation factors of 2, 3, 4, 5, 6, 2, and 3 respectively. All SE-Res2Blocks use the same dilation configuration.
\end{enumerate}

In addition, we create four models based on the ResNet34 architecture introduced in~\cite{magneto}. We enhance all our ResNet34 models by adding Squeeze-Excitation (SE) blocks~\cite{se_block} after each ResBlock component. Three additional variants are trained:

\begin{enumerate}
    \item Incorporation of a Channel-dependent Attentive Statistics (CAS) pooling layer~\cite{ecapa_tdnn}.
    \item Usage of two sub-centers per class in the AAM-softmax layer (SC-AAM).
    \item Integration of both CAS and two sub-center AAM.
\end{enumerate}

\subsection{Training protocol}

In the closed track, we train all models on the development set of the  VoxCeleb2 dataset~\cite{vox2}. For the open track we add the development part of the VoxCeleb1 \cite{vox1} dataset, the \textit{train-other-500} subset of the LibriSpeech dataset~\cite{libri} and a part of the DeepMine corpus \cite{deepmine} containing utterances from 588 Farsi speakers.
We also create 6 additional augmented copies using the MUSAN~\cite{musan} corpus (babble, noise), the RIR~\cite{rirs} (reverb) dataset and the SoX (tempo up, tempo down) and FFmpeg (compression) libraries.

All systems are trained on random crops of 2~s to prevent overfitting. The input features are 80-dimensional MFCCs with a window size of 25~ms and a frame shift of 10~ms for the ECAPA-TDNN systems. For the ResNet based systems we use 80 log Mel filter bank energies as input features. To further improve robustness of the models, we apply SpecAugment \cite{specaugment} to the log mel-spectrograms which randomly masks 0 to 5 frames in the time-domain and 0 to 8 frequency bands. Subsequently, the input features of the cropped segment are cepstral mean normalized.

The initial margin penalty of the AAM-softmax layer is set to 0.2. We also apply a weight decay of 2e-5 on the weights in the networks, except for the AAM-softmax layer, which uses a slightly higher value of 2e-4. The systems are trained using the Adam optimizer \cite{adam} with a Cyclical Learning Rate (CLR) using the \textit{triangular2} policy as described in~\cite{clr}. The maximum and minimum learning rates are set at 1e-3 and 1e-8 respectively. We use a cycle length of 130k iterations with a batch size of 128. All ECAPA-TDNN models are trained for three full cycles, the ResNet34 models use one training cycle.

\subsection{Large margin fine-tuning}

After the initial training stage we apply our proposed large margin fine-tuning strategy~\cite{voxsrc20_submission} to all models. During this fine-tuning stage, the margin of the AAM-softmax layer is increased to 0.5. SpecAugment is disabled and the length of the random crop is increased to 6~s. The CLR cycle length is decreased to 60k, with the maximum learning rate lowered to \mbox{1e-5}. No layers in the network are frozen. Finally, the sampling strategy is changed to HPM as described in~\cite{sdsvc_paper} with parameter values $S = 16$, $I = 8$ and $U = 1$. An ablation study can be found in~\cite{voxsrc20_submission}. 

We create scores for all fine-tuned systems using adaptive s-normalization~\cite{s_norm_2} with an imposter cohort size of 100. The imposter cohort consists of the average of the length-normalized utterance-based embeddings of  each training speaker. The imposter cohort size is set to 100.

\subsection{Quality-aware score calibration}

To train our calibration system we create a set of trials from the VoxCeleb2 training dataset. We want our calibration system to be robust against varying levels of duration in the trials. Subsequently, we define three types of trials: \textit{short-short}, \textit{short-long} and \textit{long-long} with \textit{short} indicating an utterance between 2 and 6 seconds and \textit{long} ranging from 6~s to the maximum length utterance in the VoxCeleb2 dataset. We include 10k trials of each type in our calibration trial set, resulting in a total of 30k trials. The amount of target and non-target trials is balanced.

We calibrate each system individually on the calibration set using logistic regression after which we calculate the mean score across all models. Subsequently, we use our proposed quality-aware calibration~\cite{voxsrc20_submission} on the mean score to make the decision thresholds of the evaluation metrics robust across varying utterance conditions. We incorporate two symmetric Quality Measure Functions (QMFs) based on the minimum and maximum quality value in the quality-aware calibration stage. Our first quality metric is the amount of speech frames detected by the Voice Activity Detector (VAD) module of the SPRAAK system~\cite{spraak}. As our second metric, we use the mean top-100 imposter score based on the inner-product of the considered embedding and the s-normalization imposter cohort.

  \begin{table}[h]
      \caption{Impact of distance metric and cohort size on imposter-based QMF on the VoxSRC-20 validation set.}
      \vspace{-5pt}
     \label{tab:ablation_qmf}
   \centering
   \scalebox{0.95}{
   \begin{tabular}{ccccc}
     \toprule
      & \textbf{Method} & \multicolumn{1}{c}{\textbf{Cohort}} & \multicolumn{1}{c}{\textbf{EER(\%)}} & \multicolumn{1}{c}{\textbf{MinDCF\textsubscript{0.01}}} \\
     \midrule
     & baseline & - & 2.89 & 0.2274 \\
     \midrule
     A & inner product & all & 2.95 & 0.2326 \\
     B & inner product & top-100 & \textbf{2.81} & \textbf{0.2257} \\
     C & cosine & all & 2.89 & 0.2274  \\
     D & cosine & top-100 & 2.89 & 0.2280 \\
     \bottomrule
   \end{tabular}}
 \vspace{-15pt}
 \end{table}
 
A detailed description of the quality-aware score calibration can be found in~\cite{voxsrc20_submission}. As an additional analysis to~\cite{voxsrc20_submission}, we compare the usage of the inner product and cosine distance during the calculation of the imposter mean against different cohort sizes in Table~\ref{tab:ablation_qmf} on our baseline ECAPA-TDNN system. Experiments \textit{A} and \textit{C} show that using the cosine distance when the imposter cohort contains all 5994 training speakers appears to be more stable than using the inner product, which even degrades performance. However, a properly chosen cohort size appears to be beneficial for the imposter mean based on the inner product, as indicated in experiment \textit{B}, for which we note a relative improvement of 3.1\% EER and 0.7\% MinDCF over the baseline system.


\subsection{Fusion system performance}

Table~\ref{tab:ablation_fusion} provides a performance overview on the Vox-SRC20 validation set of all fine-tuned systems in our final system fusion submission for the closed track. Notably, by incorporating Channel-dependent Attentive Statistics (CAS) pooling in the SE-ResNet34 system, the architecture now provides performance similar to the baseline ECAPA-TDNN system. Using a Tanh activation in the CAS layer does not seem to have a big impact. Adding a BLSTM layer does not improve performance over the baseline ECAPA-TDNN, possibly due to the SE-blocks already inserting global context information in the frame-level layers of the model. The experiment with Dynamic Dilation (DD) indicates there are multiple ways to gradually scale up the temporal context of the ECAPA-TDNN system. Interestingly, using a too large embedding dimension can degrade the performance of the ECAPA-TDNN significantly. We also notice a slight performance degradation by only using 60 log Mel filter bank energies. Using two sub-centers in the AAM-softmax (SC-AAM) layer results in additional performance gains on the ECAPA-TDNN and ResNet34 based systems. The VoxCeleb2 dataset is less curated than the VoxCeleb1 dataset and this more robust training loss might compensate for labeling errors, but it may also compensate for too strong augmentations. Incorporating SC-AAM and CAS in a SE-ResNet34 based system leads to our best single system performance.

\begin{table}[h]
\centering
    \caption{Evaluation of all fine-tuned systems in the final fusion submission of the closed track on VoxSRC-20 validation set.}
    \vspace{-5pt}
    \label{tab:ablation_fusion}
  \resizebox{0.91\textwidth}{!}{\begin{minipage}{\textwidth}
  \begin{tabular}{lccc}
    \toprule
     \textbf{Architecture} & \textbf{Variant} & \multicolumn{1}{c}{\textbf{EER(\%)}} & \multicolumn{1}{c}{\textbf{MinDCF\textsubscript{0.01}}} \\
    \midrule
     ECAPA-TDNN & Baseline & 2.89 & 0.2274 \\
     ECAPA-TDNN & Tanh in CAS & 2.86 & 0.2274 \\
     ECAPA-TDNN & BLSTM & 2.88 & 0.2360 \\
     ECAPA-TDNN & 256-dim emb. & 3.15 & 0.2578  \\
     ECAPA-TDNN & FBANK60 & 2.92 & 0.2389  \\
     ECAPA-TDNN &  SC-AAM \& DD & 2.83 & 0.2298  \\
     ResNet34 & SE-blocks & 3.03 & 0.2605 \\
     SE-ResNet34 & CAS & 2.89 & 0.2306 \\
     SE-ResNet34 &  SC-AAM & 2.98 & 0.2437 \\
     SE-ResNet34 & SC-AAM \& CAS & \textbf{2.70} & \textbf{0.2215} \\
     \midrule
     Fusion & & 2.41 & 0.1901 \\
     Fusion + QMFs & & \textbf{2.16} & \textbf{0.1795} \\
    \bottomrule
  \end{tabular}
  \end{minipage}}
\end{table}


Evaluation of our final system fusion in the closed track with large margin fine-tuning and quality-aware score calibration is given in Table~\ref{tab:ablation_voxsrc}. Large margin fine-tuning of all models results in a relative improvement of 3\% in EER and 8\% in MinDCF. Using quality-aware score calibration of the fused score with the speech duration and imposter mean QMFs resulted in an additional 8\% EER and 6\% MinDCF relative improvement.

Due to time constraints, we only trained one extra baseline ECAPA-TDNN model for the open speaker verification track. We replace the under-performing ECAPA-TDNN with 256-dimensional embeddings with this new model. Our final open fusion submission improves relative upon the closed submission with 4\% EER and 2\% MinDCF.

\begin{table}[h]
    \caption{Evaluation of the proposed fine-tuning and quality-aware calibration (QMFs) on the VoxSRC-20 test set.}
    \vspace{-5pt}
    \label{tab:ablation_voxsrc}
  \centering
  \resizebox{0.98\textwidth}{!}{\begin{minipage}{\textwidth}
  \begin{tabular}{lccc}
    \toprule
     \textbf{Systems} & \textbf{Track} & \multicolumn{1}{c}{\textbf{EER(\%)}} & \multicolumn{1}{c}{\textbf{MinDCF\textsubscript{0.05}}} \\
    \midrule
     Fusion & closed & 4.20 & 0.2052 \\
     Fusion + FT & closed & 4.06 & 0.1890 \\
     Fusion + FT + QMFs & closed & \textbf{3.73} & \textbf{0.1772}  \\
     Fusion + FT + QMFs & open & \textbf{3.58} & \textbf{0.1737}  \\
    \bottomrule
  \end{tabular}
  \end{minipage}}
\vspace{-10pt}
\end{table}

\section{Unsupervised speaker verification}

\subsection{Training without speaker labels}

Unsupervised speaker verification tackles the problem without the use of any manually created speaker identity labels (VoxSRC-20 Track 3). We define three main stages in our audio-only solution:

\begin{enumerate}
  \item Utterance-based contrastive learning
  \item Iterative clustering for pseudo-label generation
  \item Supervised training on pseudo-labels
\end{enumerate}

\subsubsection{Contrastive learning}

The contrastive learning stage~\cite{simclr,aat,semisupervised_contrastive_learning} generates positive and negative comparison pairs between all utterances in the unlabeled dataset. Through strong augmentation, two different versions of the same utterance can be made. It is clear that the speaker identity related to the processed utterance does not change, and the two generated segments correspond with a positive comparison pair. Speaker embeddings generated by the speaker verification model should be consistent within a positive pair, indicating this speaker embedding extractor is invariant to the applied augmentations. As there are no speaker labels available, it is assumed that each training utterance corresponds with a different speaker. A negative comparison pair is thus constructed by simply comparing across utterances. Embeddings within a negative pair should be different, showing that the network can differentiate between the speakers.

We rely on Momentum Contrast (MoCo)~\cite{moco} to expand the number of possible negative comparison pairs beyond the data in the mini-batch per gradient update. This MoCo technique creates a second momentum encoder, which is a copy of the original speaker embedding extractor. The momentum encoder is only used for forward passes. Instead of the back-propagation algorithm to update its model parameters, a momentum update based on the values of the original network is used during each iteration:
\begin{equation}
\theta_m \leftarrow m \theta_m + (1-m ) \theta_e
\end{equation}
with $\theta_m$ a model parameter of the momentum encoder and $\theta_e$ the corresponding model parameter in the speaker embedding extractor. The momentum $m$ controls how consistent the output will remain across multiple mini-batch update iterations.
After each parameter update, we store the momentum encoder output embeddings of the latest mini-batch in a large queue. When adding this new data to the queue, we remove the oldest momentum encoder embeddings. This way we store a large number of recent and consistent embeddings that can be used for the negative pair comparisons.

The loss function of the original speaker embedding extractor is given by:
\begin{equation}
\label{aam_softmax}
L = -\frac{1}{n} \sum^{n}_{i=1} log \frac{e^{s(\textbf{x}_i \cdot \textbf{x}^{m}_{i+})}}{e^{s(\textbf{x}_i \cdot \textbf{x}^{m}_{i+})} + \sum_{j=1}^N e^{s(\textbf{x}_i \cdot \textbf{x}^{m}_{j-})}}
\end{equation}
with $n$ indicating the batch size, $\textbf{x}$ is a length-normalized speaker embedding that is generated by the original speaker embedding extractor. $\textbf{x}^m$ refers to the normalized embeddings of the momentum encoder. Note that the embeddings $\textbf{x}_{i}$ and $\textbf{x}^{m}_{i+}$ are each generated with a different augmentation applied to the original utterance. 
The MoCo queue for negative pair comparisons contains $N$ stored embeddings $\textbf{x}^{m}_{j-}$. A large value for $N$ can be set, as the momentum encoder is not updated through back-propagation. A scaling factor $s$ is applied to increase the range of the output log-likelihoods. After the parameter update, the oldest embeddings in the MoCo queue are replaced by all $\textbf{x}^{m}_{i+}$ in the mini-batch.

\subsubsection{Iterative clustering}

The assumption that each utterance corresponds with a different speaker is most likely incorrect for most training datasets. Once a speaker embedding extractor model is trained through contrastive learning, we can use this model to characterize the speaker in each training utterance by extracting an embedding. These embeddings can be grouped by means of a clustering algorithm.

One of the most successful algorithms in speaker diarization is Agglomerative Hierarchical Clustering (AHC). However, the memory complexity does not allow to efficiently apply AHC to the complete VoxCeleb2 dataset with current hardware and software. Therefore, we rely on an extremely efficient mini-batch k-means~\cite{minibatch_kmeans, scikit-learn} cluster process to reduce the embeddings to a more manageable number of k-means cluster centers. As the contrastive learning optimized the cosine similarity between embeddings, we length-normalize the embeddings before clustering. After this initial clustering, the k-means cluster centers are grouped through AHC with Ward linkage~\cite{ward_ahc} and the cosine similarity metric.
Each utterance belonging to the same cluster of k-means centers is assigned the same speaker identity pseudo-label. We interpret these pseudo-labels as the ground truth and we train a speaker embedding extractor using the supervised AAM-softmax loss. The number of AHC output clusters can be determined by analyzing the performance of this speaker verification model on the VoxSRC validation data.
This process can be re-iterated. Given the training embeddings generated by the model trained on the pseudo-labels we can re-execute the clustering and continue training the speaker embedding extractor with the new pseudo labels. This can be repeated until performance convergence on the validation data.

\subsubsection{Robust training on pseudo-labels}

The final iteration of the clustering process generates reliable pseudo labels that can be used to train a larger speaker verification model in a supervised way. However, these pseudo labels still contain label noise and we rely on the AAM subcenter loss~\cite{subcenter} for potential robustness against these label errors. To further optimize performance, the trial scores of this large model can be fused with the scores of the final model in the preceding iterative clustering process.

\subsection{Implementation}

\subsubsection{Contrastive learning}
An ECAPA-TDNN~\cite{ecapa_tdnn} with $C=1024$ is trained on VoxCeleb2 in an unsupervised way with contrastive learning. We only use two SE-Res2Blocks to reduce the GPU memory requirement. 
The CLR settings are the same as in the initial phase of the supervised training, and we train the model during 1 cycle. The size of the MoCo queue is set to 65K, the momentum is 0.999. The scale $s$ in the contrastive loss is set to 10. We apply shuffle batchnorm~\cite{moco} on the momentum encoder to avoid non-generalizing information leakage through the batchnorm statistics. This problem is caused by the fact that all the positive pair utterances of the embedding extractor and the momentum encoder are grouped in a mini-batch corresponding with an identical group of source utterances. All other settings except the data augmentation settings are similar to the supervised learning.

As the augmentation plays a crucial role in contrastive learning we implement an online augmentation protocol. We reduce overlap between the 3.5~s random crops of the same utterance processed by the embedding extractor and the momentum encoder by sampling 5 crops and picking the pair with the least overlap. Minimizing the overlap in a more precise way, causes sub-optimal performance by frequently selecting the start and the end of the utterance during cropping. We use 37K noise segments from the balanced YouTube-8M~\cite{youtube_8m} train set for additive noise augmentation with an SNR between 5 and 15 dB. Optional reverb~\cite{rirs} is applied with a chance of 75\%. Contrary to our expectations, we noticed slight performance gains by applying reverb after the additive noise augmentation. No SpecAugment is applied. 

\subsubsection{Iterative clustering}

To guarantee high quality embeddings, we extract training embeddings with the model trained through contrastive loss from the clean full-length VoxCeleb2 utterances. The normalized embeddings are grouped to 50K clusters by Euclidean mini-batch k-means clustering with a batch size of 10K. Random initialization of the k-means algorithm is used.
The assignment of 50K cluster centers makes AHC computationally feasible in the next stage. The AHC clustering is realized using the fastcluster package~\cite{fastcluster}.
We assume that it is realistic to determine the optimal number of AHC clusters in VoxCeleb2 in steps of 2.5K extra clusters. Evaluation on the VoxSRC-20 validation set of the ECAPA-TDNN trained with the AAM-sofmax loss on pseudo-labels, showed that 7.5K clusters delivered optimal results. This analysis is shown in Figure~\ref{fig:ahc_performance}.

The data preprocessing in this stage is identical to the supervised setting, and a standard ECAPA-TDNN with three SE-Res2Blocks is used. To speed up the training process we reduce the CLR cycle length to 60K. The new training embeddings are re-clustered after one CLR training cycle. The maximum CLR learning rate does not decay per cycle as the system should be able to adapt to the improved pseudo-labels. The assigned cluster labels are permuted after each iteration. To assure that we can continue training with the same model, we replace the AAM prototypes by the mean vector of all normalized embeddings assigned to the corresponding cluster. Performance converged after 7 iterations.

\subsubsection{Robust training on pseudo-labels}

The pseudo-labels of the final clustering iteration are used to train a large ECAPA-TDNN with $C=2048$ with dynamic dilation within 4 SE-Res2Blocks and subcenter AAM with 2 centers per cluster. The CLR schedule is restored to its original setting. After 3 cycles the model is fine-tuned with the large margin setting. We do not apply any score normalization or calibration. These post-processing steps could be enabled by generating target and non-target trials from the pseudo-labels. The output scores of this system can be fused with the final model of the iterative clustering process by taking the average of the scores. This is possible without any calibration as the systems have been trained on the same input data with a similar AAM loss function.

\subsection{Results}

\begin{figure}[h]
  \centering
  \includegraphics[scale=0.275]{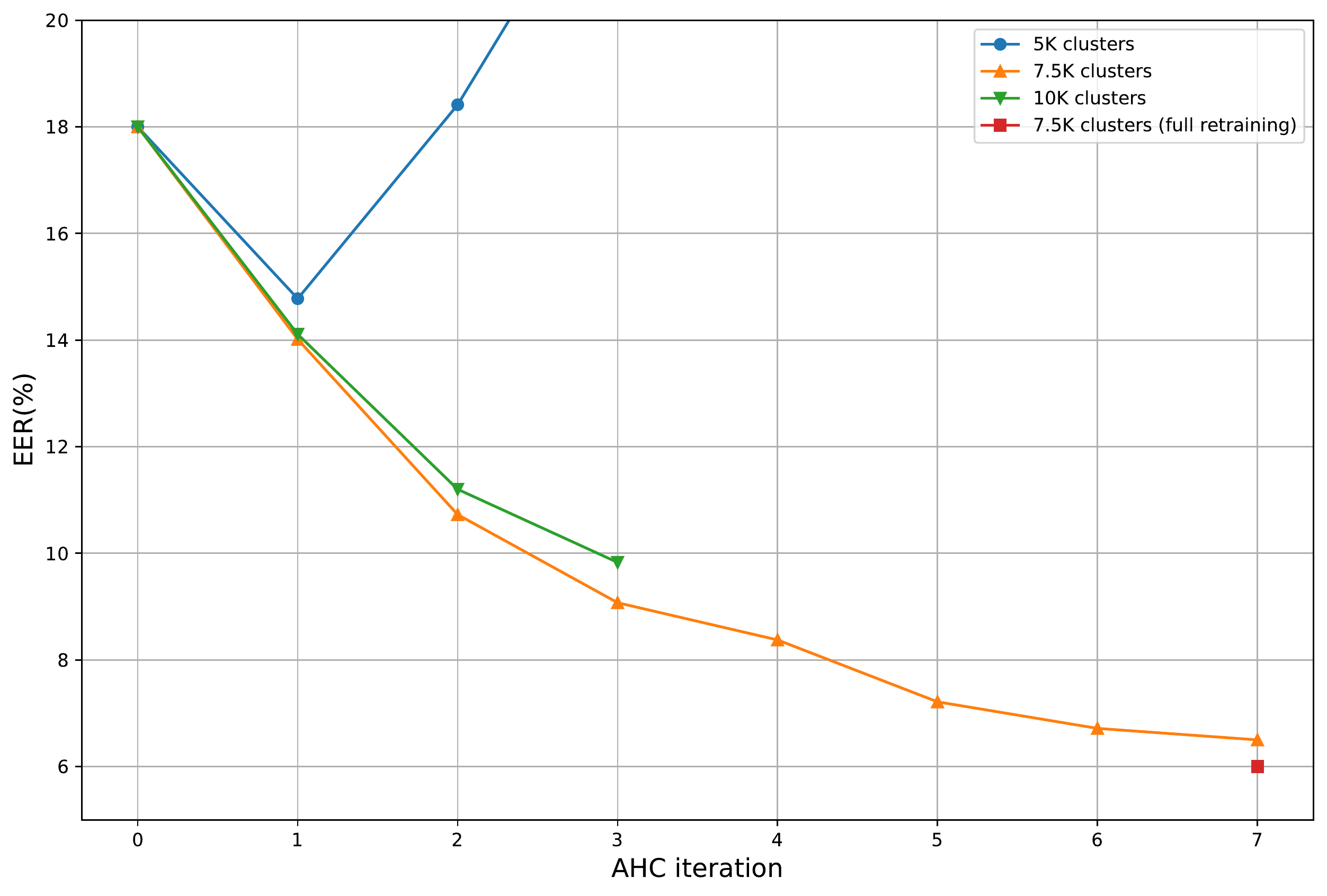}
  \setlength{\belowcaptionskip}{-10pt}
  \caption{Performance of iterative clustering on the VoxSRC-20 validation set.}
  \label{fig:ahc_performance}
\end{figure}

Performance of the unsupervised speaker verification approach is shown in Figure~\ref{fig:ahc_performance}. Contrastive learning achieves an EER of 18.0\% on the VoxSRC-20 validation set and an EER of 7.3\% on the original VoxCeleb1 test set. Iterative grouping through clustering improves the results if the number of clusters is chosen appropriately. Setting the number of clusters too low will result in performance divergence as shown in the case of 5K clusters. Constructing 7.5K clusters during the iterative process is more optimal than generating 10K clusters. In the case of 7.5K clusters the EER gradually improves to 6.5\%, which is a relative improvement of 64\% compared to the initial model. This final model achieves an EER of 2.1\% on the original VoxCeleb1 test set. Training and large margin fine-tuning of a larger ECAPA-TDNN system trained with 2-subcenter AAM on the final pseudo-labels produces an EER of 6\% on the VoxSRC-20 validation set.

The converged iterative clustering algorithm with 7.5K clusters initialized by contrastive learning achieves an EER of 7.7\% on the VoxSRC-20 test set. Score averaging with the larger fine-tuned ECAPA-TDNN system results in a final submission score of 7.2\% EER.




\bibliographystyle{IEEEtran}

\bibliography{mybib}


\end{document}